%% file: bare_jrnl.tex
\pdfoutput=1

\documentclass[journal]{IEEEtran}
\usepackage{bbold,bbm,bm, amssymb, graphicx, amsthm, caption, subcaption,multicol,amsmath}
\usepackage{tabu}
\usepackage{diagbox}
\usepackage{multirow}
\usepackage{booktabs}
\usepackage{algorithm}
\usepackage[noend]{algpseudocode}
\usepackage[dvipsnames]{xcolor}
\usepackage{hyperref}
\usepackage[numbers,sort&compress]{natbib}
\usepackage{url}
\usepackage{footnote}
\usepackage{perpage}
\makeatletter
\def\blfootnote{\xdef\@thefnmark{}\@footnotetext}
\makeatother
\usepackage{enumitem}

\theoremstyle{plain}
\newtheorem{thm}{Theorem}

\theoremstyle{remark}

\newcommand{\bomega}{\bm{\omega}}

\newcommand{\bomegahat}{\hat{\bm{\omega}}}
\newcommand{\htilde}{\tilde{h}}
\newcommand{\ptilde}{\tilde{p}}
 
\newcommand{\argmin}{\mathop{\mathrm{argmin}}}

\newcommand{\bH}{\mathbf{H}}

\newcommand{\bx}{\mathbf{x}}

\newcommand{\bh}{\mathbf{h}}

\begin{document}

\title{Interference management and capacity analysis for mm-wave picocells in urban canyons}

\author{Zhinus~Marzi 
        and~Upamanyu~Madhow,~\IEEEmembership{Fellow,~IEEE}\\
        \texttt{\{zhinus\_marzi, madhow\}@ucsb.edu}
\thanks{Copyright (c) 2019 IEEE. Personal use of this material is permitted. However, permission to use this material for any other purposes must be obtained from the IEEE by sending a request to pubs-permissions@ieee.org} \\ 
\thanks{Z. Marzi and U. Madhow are with the Department
of Electrical and Computer Engineering, University of California Santa Barbara, Santa Barbara,
CA}}
        \maketitle

\begin{abstract}

Millimeter (mm) wave picocellular networks are a promising approach for delivering the 1000-fold capacity increase required to keep up with projected demand for wireless data: the available bandwidth is orders of magnitude larger than that in existing cellular systems, and the small carrier wavelength enables the realization of highly directive antenna arrays in compact form factor, thus drastically increasing spatial reuse. In this paper, we carry out an interference analysis for mm-wave picocells in an urban canyon with a dense deployment of base stations.  Each base station sector can serve multiple simultaneous users, which implies that both intra- and inter-cell interference must be managed.  We propose a \textit{cross-layer} approach to interference management based on (i) suppressing interference at the physical layer and (ii) managing the residual interference at the medium access control layer. We provide an estimate of network capacity, and establish that 1000-fold increase relative to conventional LTE cellular networks is indeed feasible.


\end{abstract}


\IEEEpeerreviewmaketitle


\section{Introduction}
\input{intro_UM}

\section{System model} \label{sec:sys}
\input{Sys_model}

\section{Inter-cell interference} \label{sec:inter}
\input{Inter_cell}

\section{Intra-cell interference} \label{sec:intra}
\input{intra_cell3}

\section{Capacity estimation} \label{sec:capacity}
\input{Capacity2}

\section{Discussion and conclusions}
\input{Conclusion_UM}

\section*{Acknowledgment}
This work was supported in part by the National Science
Foundation under grant CNS-1518812.

\bibliographystyle{ieeetr}
\bibliography{bare_jrnl.bbl}

\vspace{70mm}

\begin{IEEEbiography}[{\includegraphics[width=1.1in,height=1.35in,keepaspectratio]{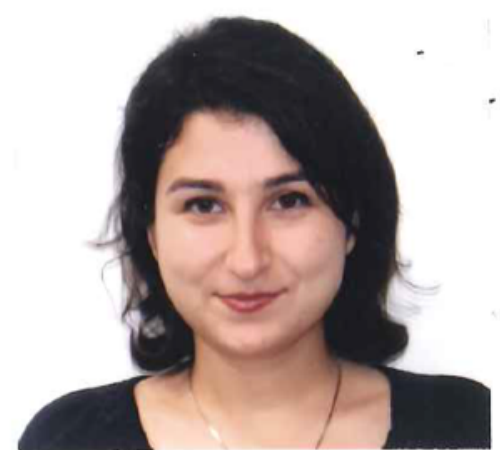}}]{Zhinus Marzi}
is currently pursuing the Ph.D. degree with the department of electrical and computer engineering at the University of California Santa Barbara (UCSB). She has received her B.Sc. and M.Sc. in electrical engineering from the Iran University of Science and Technology and Sharif University of Technology, Tehran, Iran, in 2010 and 2012 respectively. She is interested in wireless communication and signal processing and her current focus is on millimeter wave communications.
\end{IEEEbiography}

\vspace{-130mm}

\begin{IEEEbiography}[{\includegraphics[width=1in,height=1.25in,clip,keepaspectratio]{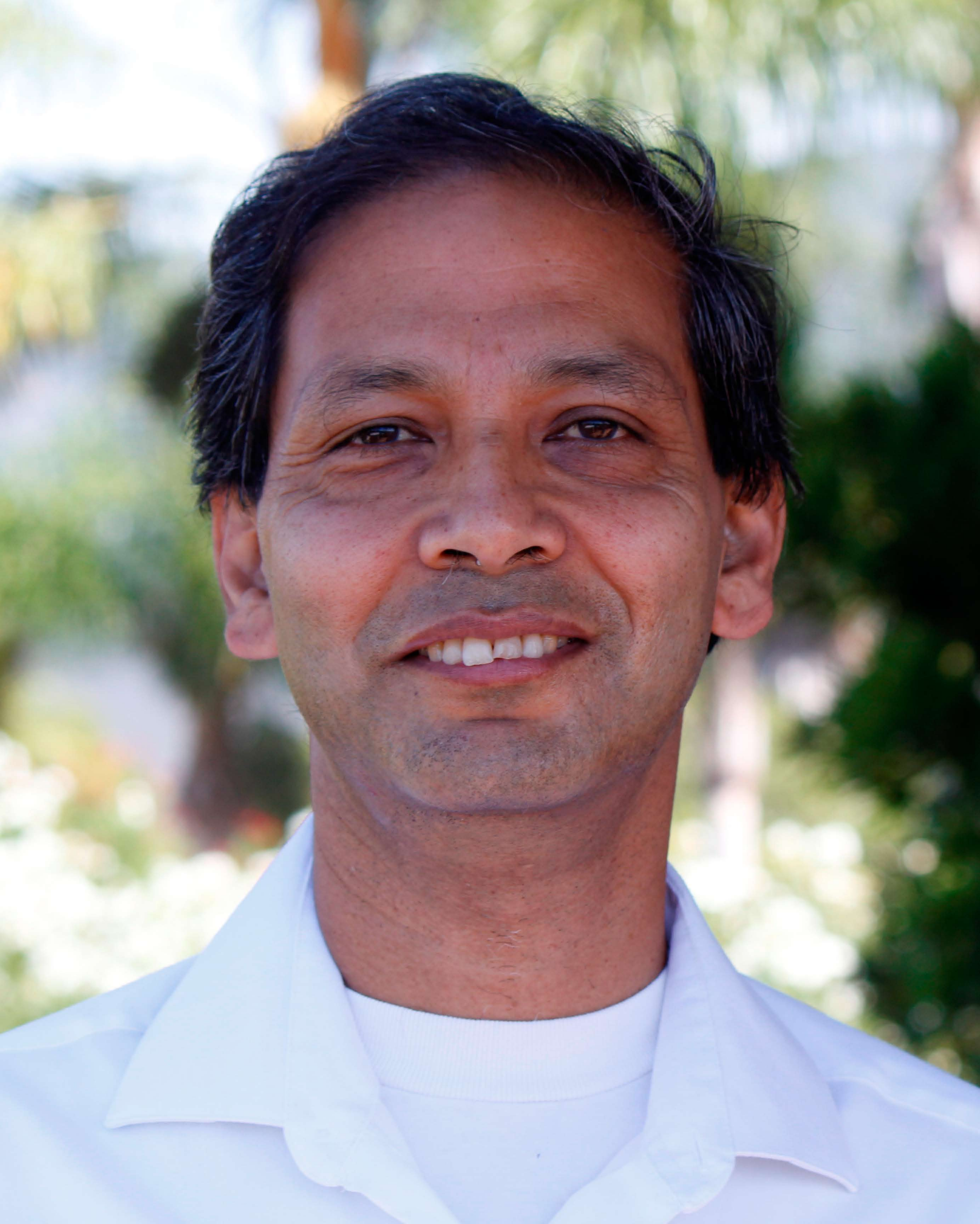}}]{Upamanyu Madhow}
is Professor of Electrical and Computer Engineering
at the University of California, Santa Barbara. His research interests broadly span
communications, signal processing and networking, with current emphasis on
millimeter wave communication, and  on distributed and bio-inspired approaches to networking and inference.
He received his bachelor's degree in electrical engineering from the Indian Institute of Technology, Kanpur, in 1985, and his Ph.D. degree
in electrical engineering from the University of Illinois,
Urbana-Champaign in 1990. He has worked as a research scientist at
Bell Communications Research, Morristown, NJ, and as a faculty at the
University of Illinois, Urbana-Champaign. Dr. Madhow is a recipient of the 1996 NSF CAREER award, and co-recipient of the 2012 
IEEE Marconi prize paper award in wireless communications. He has served as Associate Editor for the IEEE Transactions on
Communications, the IEEE Transactions on Information Theory, and the 
IEEE Transactions on Information Forensics and Security. He is 
the author of two textbooks published by Cambridge University Press, Fundamentals of Digital Communication (2008) and Introduction to Communication Systems (2014).
\end{IEEEbiography}

\end{document}

%% file: intro_UM.tex

Recent years have seen an explosion in cellular data demand due to bandwidth-hungry multimedia applications. This is projected to require a \textit{1000-fold} capacity gain by 2020 \cite{chih20165g}.
In response to this demand, both industrial and academic communities have converged upon the mm-wave frequency band (30-300 GHz) as the next frontier for cellular communication \cite{5G_industry,rappaport_survey_2014}.
This is because of two major reasons. 
First, this frequency band offers an enormous amount of bandwidth \footnote{For example, FCC has allocated 14 GHz of contiguous unlicensed spectrum in the 60 GHz} compared to existing cellular networks. Second, the short wavelength at this band ($\leq$ 10 mm) means that electronically large antenna arrays can be made physically small\footnote{For example at 60 GHz, an 8 $\times$ 8 array occupies an area of less than a square inch, while a 32 $\times$ 32 array fits within 10 square inches.}, enabling highly directive links. We can exploit the resulting reduction in interference by a dense deployment base stations yielding a drastic increase in spatial reuse relative to existing systems.

\begin{figure}[t]
 \begin{center}
    \includegraphics[width=3.5in,height=1.5in]{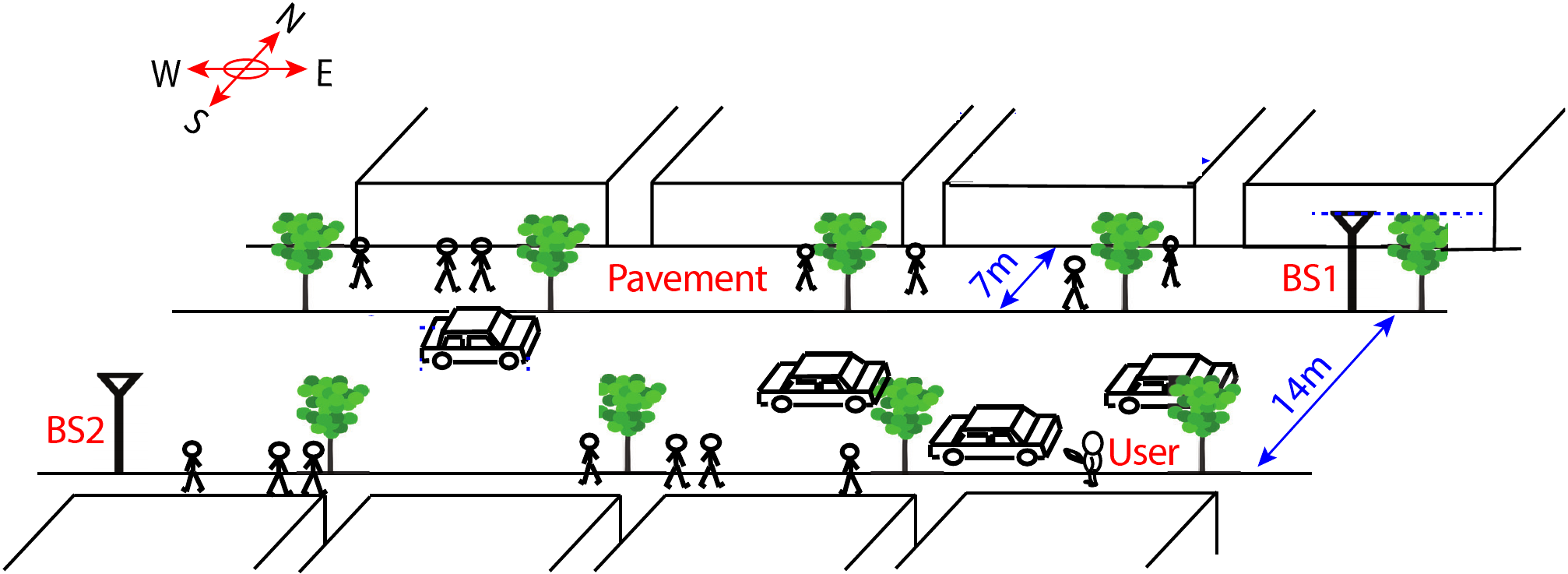}
    \vspace{1mm}
\caption{Picocellular network deployed along an urban canyon}
\vspace{-8mm}
\label{fig:street}
\end{center}
\end{figure} 


There is a growing body of research on the feasibility of mm-wave small cells in terms of link budget and channel modeling \cite{rappaport2013millimeter,zhu2014demystifying,rappaport2013broadband,zhang2010channel,karjalainen2014challenges}.  There is also a recognition that the problem of beam discovery
and user tracking is a particularly important one \cite{marzi2016compressive,ramasamy2012compressive2,ramasamy2012compressive1,rasekh2017noncoherent,Hassanieh:2018:FMW:3230543.3230581}:  mobile users must be accurately tracked in order to form highly directive beams, and the relative ease of blockage of mm waves implies that an inventory of multiple feasible paths to a given
user must be maintained in order to facilitate switching in the event of blockage. Providing an adequate backhaul for mm-wave picocells
is another challenge, with mm-wave backhaul (possibly using a band different from that used for the access link) as one viable option \cite{islam2017integrated,kulkarni2018many,rasekh2015interference,rasekh2018joint}.
In short, there are many challenges that must be addressed in order to realize the system concept driving the work reported here.
In this paper, however, we sidestep these issues, assuming that such challenges will be eventually surmounted, and focus on estimating
the capacity of the resulting system. In order to do so, we must characterize the interference in such a system, and provide sensible interference
management strategies that are tailored to the unique characteristics and geometry of the system.

%

While the system design concepts presented here are of rather general applicability, our numerical results are for a particular setting that we feel has great promise, as also discussed in some of our prior publications \cite{marzi2015interference,marzi2016compressive,zhu2014demystifying,ramasamy2012compressive2}. We propose to employ the 60 GHz unlicensed band for base station to mobile communication in outdoor picocells: More specifically, we consider picocellular base stations deployed on lampposts on each side of the street along an urban canyon (e.g. a typical street in New York City), as depicted in Figure \ref{fig:street}. (discussed further in Section \ref{sec:sys}).  Each base station ``face'', or sector, could potentially support multiple simultaneous users.
We currently assume that this is accomplished by employing multiple subarrays, each capable of RF beamforming to a different user.  Alternatively, if and when digital beamforming becomes feasible for large mm-wave arrays, a single array could simultaneously form beams towards multiple users.

\subsection{Contributions}

Prior work at lower carrier frequencies shows that interference becomes
a fundamental limiting factor in picocellular settings \cite{ramasamy_ISIT13}. However, as we show here, the narrow beams synthesized using large arrays at 60 GHz alleviate this problem. Here is a brief overview of our roadmap to estimate the capacity gain of mm-wave picocellular networks.

We characterize \textit{inter-cell} interference, using an analysis accounting for the geometry of the urban canyon. The approach involves
studying the interference caused by main beam and sidelobes separately, since they have distinct characteristics. This is largely a summary
of work reported in our previous conference paper \cite{marzi2015interference}. While this prior work considers only one subarray per base station face, 
it extends naturally to the multiple subarray scenario considered here.

The key challenge addressed in this paper is to quantify the gain in spatial reuse by employing multiple subarrays per base station face.
The effect of additional inter-cell interference caused by increase in the aggregate number of transmitters in the system is characterized by
adapting our prior analysis in \cite{marzi2015interference}.  However, the characterization and management of
the \textit{intra-cell interference} originating from the other transmitting subarrays on the same base station is challenging, and is the
main thrust of this paper.




Specifically, we propose a \textit{cross-layer} approach to deal with the intra-cell interference in which we combine techniques from two broad areas that have been studied in the literature: (a) downlink linear precoding and power control \cite{visotsky1999optimum,wiesel2006linear,ulukus1998adaptive,rashid1998transmit,rashid1998joint} (b) powerful optimization approaches recently developed for network-level resource allocation \cite{zhuang2014traffic,zhuang2018large}. Here is a brief description of our two-step method: 
\begin{enumerate}
\item Given that a resource block is assigned to a pre-defined set of users, we develop a PHY-layer building block which employs an optimal linear method (i.e.,\ LMMSE) for beamforming and power allocation to supress the LoS intra-cell interference among them.
\item We then incorporate the PHY-layer block in designing the MAC-layer protocol which solves an optimization problem to determine the set of active users on each resource block. 
\end{enumerate}

Finally, we evaluate our proposed scheme via simulations of picocells along an urban canyon, taking both inter- and intra-cell interference into account. We then compute the overall capacity per square kilometer for a typical region in Manhattan area, and demonstrate that dense mm-wave picocellular networks can actually deliver the promised 1000-fold capacity increase over the today's cellular networks.

\subsection{Related work}

There are a number of prior papers that investigate the capacity of mm-wave networks in various architectures. Among those \cite{akdeniz2014millimeter,abouelseoud2013system,bai2015coverage,akoum2012coverage,kim2014system} study outdoor cellular network architecture.

Authors in \cite{rangan2014millimeter,akdeniz2014millimeter} show that spectral efficiency in mm-wave cellular systems can reach that of state-of-the-art LTE systems by employing high directional antennas. They consider a 1-GHz bandwidth time-division duplex (TDD) for mm-wave system which could easily provide a 20-fold increase in average cell throughput in comparison to a 20+20-MHz LTE system. Hence the capacity gain essentially comes from the bandwidth gain, and in contrast to the present work, they do not explore the spectral efficiency improvement due to highly directional antennas. Moreover, \cite{rangan2014millimeter} considers hexagonally shaped cells where the base stations are also placed randomly, as opposed to our more structured scenario of regularly placed base stations in an urban canyon. 

Similarly, \cite{abouelseoud2013system} conducts system level simulations of the 60 GHz band for outdoor scenarios like college campuses and urban environments in order to evaluate the capacity of mm-wave networks. Despite their use of large 20$\times$20 antenna arrays (compared to 8$\times$8 in this paper), their overall capacity estimate is much smaller than ours (400 Gbps/km$^2$ vs. 2.7 Tbps/km$^2$ even for our least sparse scenario). This is because \cite{abouelseoud2013system} does not employ any interference suppression schemes (other than conventional beamforming) or opportunistic resource allocation strategies. They instead apply a round-robin scheme that fails to adapt to the spatial diversity of users to handle interference. This prohibits dense deployment of base stations, resulting in capacity saturation at a much lower level compared to ours.

Coverage and attainable data rates in outdoor mm-wave networks are also investigated in \cite{bai2015coverage}, which uses stochastic geometry models, with base stations, users and obstacles placed in the 2-D plane according to Poisson point processes. This is different from our structured 3-D model with regular base station placement. Following that, \cite{akoum2012coverage} considers a detailed mm-wave channel model and considering the same stochastic model, they  predict 50-fold capacity gain while keeping the same coverage at mm-wave. They do not exploit mm-wave large antenna arrays to suppress interference while employing zero-forcing in microwave scenario and hence the capacity gain for mm-wave band is solely due to larger bandwidth (1GHz vs. 20 MHz).  

There are a few other papers which study the mm-wave networks capacity in other architectures. For example, \cite{qiao2015enabling} study the enabling of device-to-device (D2D) mm-wave links coexisting with 4G cellular networks. They establish that the resource sharing optimization problem of this scenario is hard to solve (integer nonscalable optimization problem), and propose a heuristic approch which avoids the LoS interference. This leads to higher aggregate capacity (compared to 4G cellular networks) through a larger number of concurrent transmissions. Authors in \cite{kim2014system}, conduct extensive simulation for a complicated urban environment in Korea. They investigate a multilevel topology through wireless backhaul link and examined the effects of antenna configuration (arrangment, titling angle and spacing) on coverage and capacity. 

The present paper differs from the preceding body of work in two main aspects. Firstly, capacity and interference analysis for the urban canyon model (which is well matched to big cities where there is greatest demand for mobile capacity) and structured placement of base stations 
has not been considered in prior work, except for our own preliminary results reported in \cite{marzi2015interference}.
Secondly, we explore opportunities to improve spectral efficiency in mm-wave networks while capacity gains attained in previous works are solely due to the larger bandwidth of mm-wave band. The main contribution of this work is to propose and evaluate a cross-layer approach, by utilizing large antenna arrays to suppress interference, and employing novel scheduling approaches to handle residual interference induced in dense deployment of base stations.

The work with the closest perspective to ours is \cite{yiu2009empirical}, which evaluates the mm-wave capacity for WLANs. They consider a single room, with a 60 GHz access point in the center of the ceiling and users uniformly distributed in the room. They employ a \textit{heuristic static} predefined space time division multiple access (STDMA) algorithm that separates users in either space or time domain. Specifically, they first partition the room into less overlapping regions (considering the level of interference the access point introduce to other partitions when serving a user in a particular partition) and then define which partitions could be covered simultaneously while their mutual interference is attenuated by employing nullforming. However, their static approach for fixed and simple indoor environments is not directly applicable to the more dynamic and complicated scenarios like urban canyons. Similarly, \cite{cai2007spatial} study the achievable spatial multiplexing gain in mm-wave WPAN networks. They define Exclusive Region (ER) for each of the flows based on a simplified model of the antenna pattern in a 2-D scenario and concurrent transmission are only favorable when they are outside each others ERs. Their approach is also hard to extend to 3-D scenarios like an urban canyon and seems to be more conservative in the sense of allowing concurrent flows compared to our dynamical cross-layer approach.

As mentioned, the present paper builds upon our previous work \cite{marzi2015interference}, which focused on inter-cell interference.
In this paper, we push the limits of spatial reuse by serving multiple users inside the cell.

%% file: Sys_model.tex

In this paper, we consider street canyons where base stations are
placed in a zig-zag pattern, such that immediate neighbors are on opposite sides of the street. Each base stations has two sets of antenna arrays placed on opposite faces, aligned such that one set faces east and the other faces west.  

Figure \ref{fig:street} depicts a canyon segment between two neighboring base stations BS1 and BS2, separated by distance d. We term such a canyon segment a \textit{picocell} of width d. Each user in the picocell could be served by either an eastward-facing antenna of BS2 or a westward-facing antenna of BS1. Thus, each picocell is covered by two sets of arrays, each belonging to a different BS.     

We now describe the channel model accounting for the sparse multipath characteristic of this band \cite{correia1996wideband,lovnes1994channel}. Sparse mm-wave channels can accurately be estimated by efficient algorithms proposed in literature \cite{marzi2016compressive}. We assume that the channel knowledge is available at both the base station and mobile users. Consider a base station bearing K antenna arrays on each face. Note that link distances are large enough that all transmitters installed on a face could be approximated as co-located from the users point of view. Therefore, the channel matrix from any of these K transmitters on each face to the q-th user is the same and denoted by $\bH_{q}$. 
Channel matrix $\bH_{q}$ is of size M $\times$ N where M is the antenna size of the mobile user and N that of the transmitter and is characterized by the path loss and spatial frequencies between any of the K transmitters and the q-th mobile user. We assume $\bH_{q}$ is known to all K transmitters as well as the q-th mobile user.


%% file: Inter_cell.tex

In this section we review our analysis and draw the main conclusions of our previous work on characterizing intercell interference \cite{marzi2015interference}. We define intercell interference as the interference induced by the transmitters on other basestations. To assess the intercell interference we have made two simplifying assumptions (a) we ignore interference across parallel urban canyons, as well as interference
which might leak from cross streets; (b) we do not consider potential reflections from horizontal ledges.  However, while more detailed modeling are needed to refine the interference and capacity estimates provided here to account for such effects, we expect the qualitative conclusions to remain unchanged.

We have investigated the inter-cell interference caused by the main lobe and side lobes separately, for they have different characteristics.
Since we consider a large number of antenna elements, the main beam is narrow and is well modeled by a single ray while side lobes are much weaker, but their directions are difficult to predict, hence we must be more careful in bounding their effect.
In the following subsections we will elaborate this by reviewing two theorems from our previous work, the proof of which can be found in \cite{marzi2015interference}.

\subsection{Main lobe interference} 

We consider transmitters with a large number of elements forming a pencil beam towards the desired user.
This ``desired'' beam can be along the LoS, or it can be a single bounce from a wall or the ground (e.g., when steering around an obstacle
blocking the LoS). Given the highly directive nature of the beam and the limited diffraction at small wavelengths \cite{5450459} we can use ray tracing to understand the interference such a beam creates for neighboring basestations. 

\begin{figure}[htbp]
 \begin{center}
    \includegraphics[width=3.5in,height=1.5in]{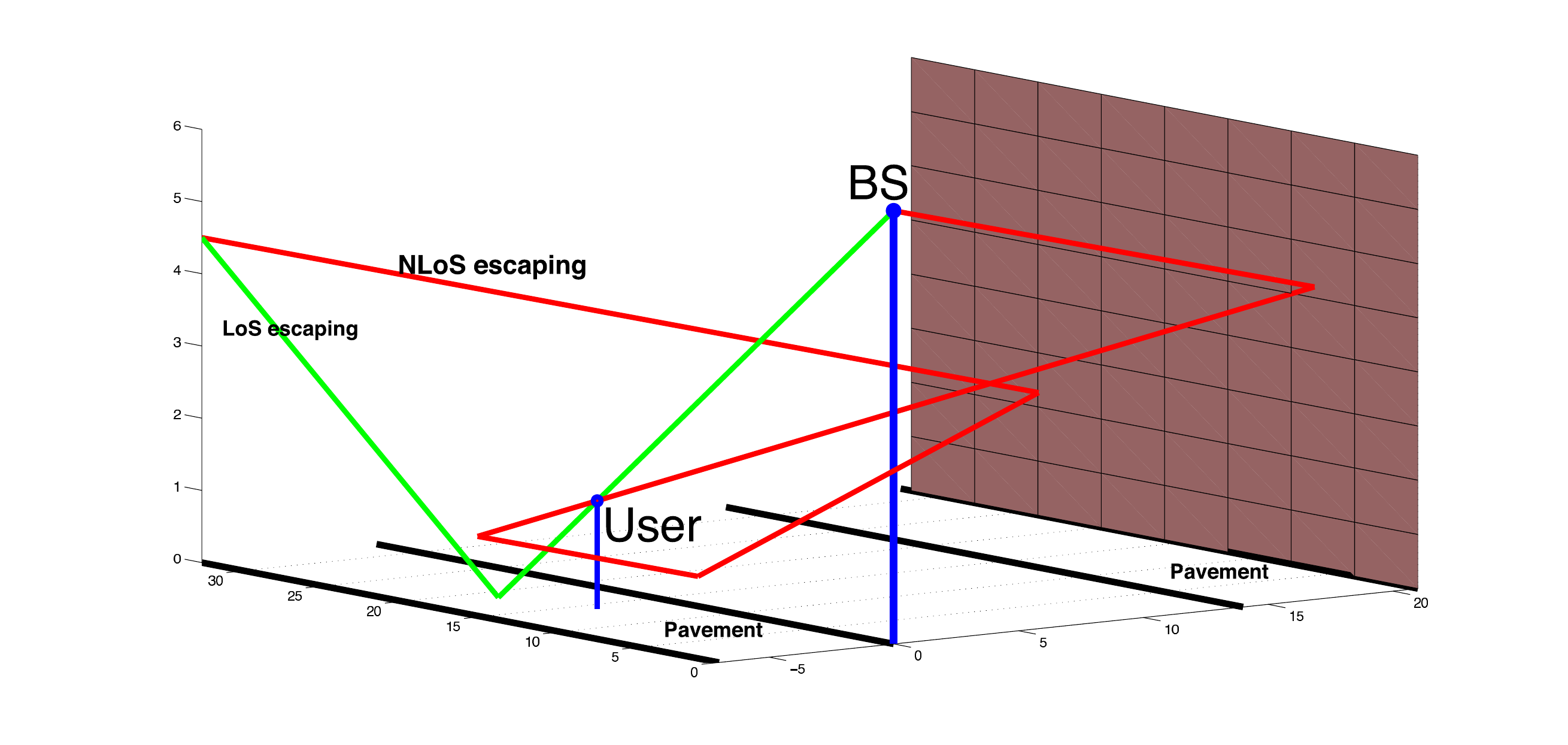}
\caption{Mainlobe will escape to sky after a few bounces}
\label{fig:ray}
\end{center}
\end{figure}

We previousely demonstrated that the main beam will escape to the sky after a few bounces (Figure \ref{fig:ray}), assuming that we can ignore
the effect of potential reflections from horizontal ledges. Specifically, in Theorem \ref{thm:Th1} bounds the number of neighboring cells that are affected by main beam's interference assuming that each face only creates interference in the direction it is facing.

\begin{thm}
\label{thm:Th1}
The maximum range over which the main beam can create interference is bounded by $\frac{H_{BS}+h_{max}}{H_{BS}-h_{max}} d$. 
Thus, the main beam from a face creates interference for at most $N_{max} =  \lceil \frac{H_{BS}+h_{max}}{H_{BS}-h_{max}} \rceil$ adjacent BSs
in the direction it is facing. We denote by $h_{max}$ the maximum height of users, by $H_{BS}$ the height of a basestation, and by $d$ the width of a picocell shared among two opposite facing antennas on adjacent basestations.
\end{thm}

For typical values of $H_{BS}=6m$ and $h_{max}=2m$ employed in our simulations, Theorem \ref{thm:Th1} implies that the main beam interferes with two adjacent basestations in the direction of the face producing the beam.

\subsection{Sidelobe interference}

While the main beam points towards a user inside the picocell, the direction of emission of sidelobes is highly variable, hence it is not possible to limit side lobe interference to a finite number of adjacent picocells. However, as shown in \cite{marzi2015interference}
the cumulative sidelobe interference seen within a given picocell
is bounded (to a relatively small value), because of the geometric decay (with distance) of the strength of the interference from a distant picocell
caused by oxygen absorption and reflection losses, along with the quadratic decay due to path loss.
Specifically, for a user served by BS$_0$, theorem \ref{thm:Th2} has quantified interference from basestations [c,$\infty$) and (-$\infty$,-c]. (c $\ge$ 0).

Denote by $P$ the smallest received power over the desired link, which is given by
\begin{equation}
P=P_{Tx}G_{Tx}G_{Rx} (\frac{\lambda}{4 \pi L_{max}})^2 e^{-\beta L_{max}}
\end{equation}
where $P_{Tx}$,$G_{Tx}$ and $G_{Rx}$ are the transmitter power and the gains of Tx and Rx antenna arrays, respectively.
The parameters $\lambda $, $\beta $ and $L_{max}$ denote, respectively, the wavelength, oxygen absorption coefficient (16 dB/km) and maximum length of a link inside a picocell. 

\begin{thm}
\label{thm:Th2}
For a user in cell 0, the sidelobe interference due to the BSs $[c,\infty)$ and $(-\infty,c]$ is bounded by $\alpha_c P$, where $P$ is the smallest received power over the desired link.
\begin{equation}
\label{eqn:alphak}
\alpha_c=\frac{\sum\limits_{n=c}^\infty I_{n}+\sum\limits_{n=-\infty}^{-c} I_{n}}P
\end{equation}
where $\alpha_c$ decays geometrically with $c$.
\end{thm}

In brief, by Theorem \ref{thm:Th1}, if we wish to avoid main beam interference, then 
$\lceil \frac{H_{BS}+h_{max}}{H_{BS}-h_{max}} \rceil$ adjacent BSs have to coordinate. For $H_{BS}=6m$ and $h_{max}=2m$, this means
that every 3 adjacent BSs have to coordinate. Suppose, for example, that we orthogonalize transmissions among such sets of
3 basestations (i.e., with a frequency reuse of 3).  
Moreover, from the computations associated with Theorem \ref{thm:Th2}
shown in \cite{marzi2015interference}, the cumulative interference caused by sidelobes from basestations beyond this set (c $\geq$ 3) is at least 40dB weaker than the desired received power. Thus, a frequency reuse of 3 leads to very large SINR. 
In our simulations, we (somewhat arbitrarily) set $s_M$ = 6 bps/Hz,
corresponding to uncoded 64-QAM as the highest supported spectral efficiency. \footnote{Such large constellations may be a stretch
with today’s hardware, given the phase noise in mm wave
radios and the difficulty of high-precision digitization at large
bandwidths, but we hope that such hardware limitations would
be overcome in the future.} The spectral efficiency is then
shown to be bounded only by hardware considerations \cite{marzi2015interference}.
However, given the interference reduction due to narrow beams, such orthogonalization is wasteful and much larger network capacity can be obtained by imposing small coordination among base stations while keeping spatial reuse one.

%% file: intra_cell3.tex

In addition to cell densification, one can attain further spatial reuse \textit{within} the cell by increasing the number of subarrays on each base station. However, this benefit comes with the pitfall of \textit{intra-cell inteference}, i.e., when a transmitter interferes with receivers in the \textit{same} cell that it does not target. This could significantly reduce the spectral efficiency of spatially correlated users.

\begin{figure}[h]
\centering
\vspace{-2mm}
    {\includegraphics[width=0.8\columnwidth,height=1.8in]{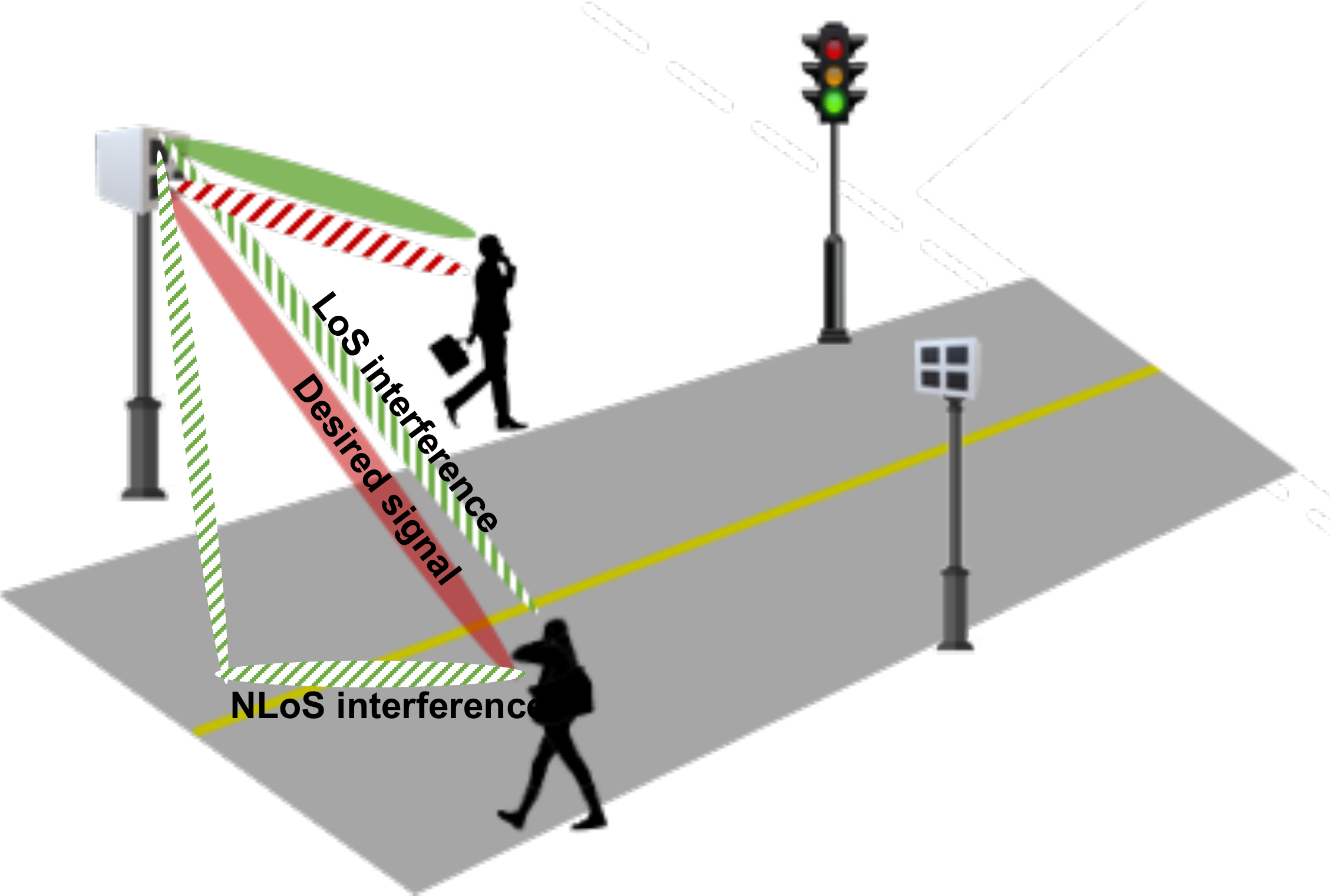}}
    \vspace{2mm}
\caption{Multiple subarrays placed on each face of a basestation which leads to intra-cell interference.}    
\label{fig:subarray}
\end{figure}

In this section, we consider K subarrays placed on each face of a basestation (Figure \ref{fig:subarray}). We first characterize intra-cell interference in our system model and then propose a \textit{cross-layer} approach to deal with it. To this end, we combined techniques from two broad areas that have been studied in the literature: (a) downlink linear precoding and power control \cite{visotsky1999optimum,wiesel2006linear,ulukus1998adaptive,rashid1998transmit,rashid1998joint} (b) powerful optimization approaches recently developed for network-level resource allocation \cite{zhuang2014traffic,zhuang2018large}. Here is a brief description of our two-step method: 

\begin{enumerate}
\item Given that a resource block is assigned to a pre-defined set of users, we develop a building block at the PHY-layer, which employs an optimal linear method (i.e.,\ LMMSE) for beamforming and power allocation to suppress the LoS intra-cell interference among them.
\item We then incorporate the PHY-layer block in designing the MAC-layer protocol, which determines the set of active users on each of the resource blocks.
\end{enumerate}

We then evaluate our proposed scheme via comprehensive simulations of picocells along an urban canyon in which both inter- and intra-cell interference are taken into account. Our simulation results demonstrate that, as we shrink cells (down to the cell width of 20m), users' spectral efficiency is mostly ( $\geq$ 97\% ) limited by the hardware limitations. A quick-glance comparison with our previous results \cite{marzi2015interference}, indicates that we are able to increase the capacity by a factor of K (at least for small number of subarrays per face i.e., K=2) in small cells. Larger picocells are more prone to interference and do not enjoy multiple subarray architecture as much, yet our proposed scheme provides users with sufficient spectral efficiency to attain large network capacity gain. Lastly, we computed the overall capacity per square kilometer for a typical region in Manhattan area and demonstrated that dense mm-wave picocellular networks can actually deliver the promised 1000-fold capacity increase over the conventional LTE networks.

\subsection{Intra-cell Interference Characterization}

\begin{figure}[h]
\centering
    {\includegraphics[width=1\columnwidth]{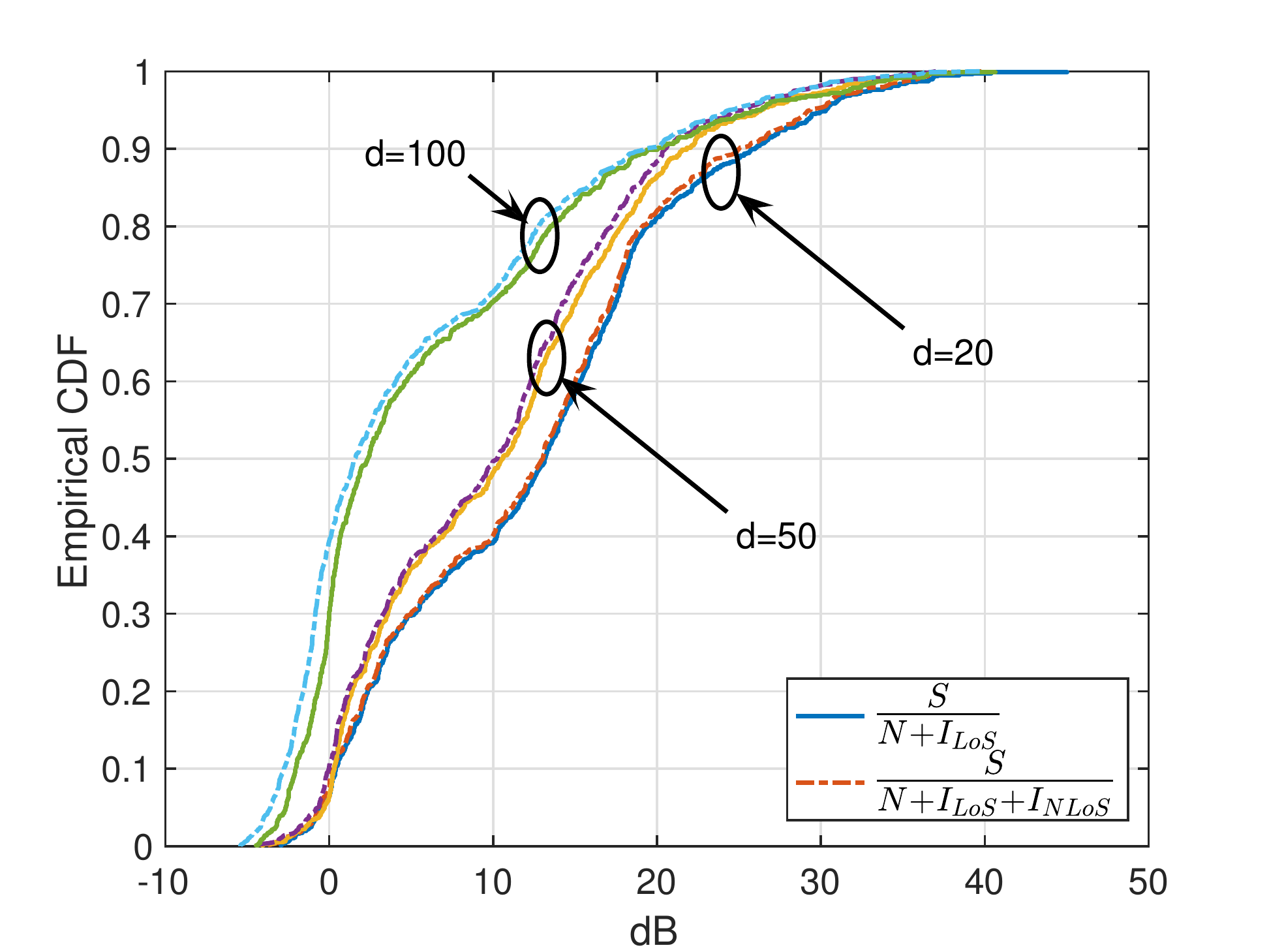}}
\caption[LoS and NLoS components of intracell interference]{CDF of signal to intracellular interference for different cell widths}  
\label{fig:LoS_NLoS}
\end{figure}

Similar to the inter-cell case, intra-cell interference is composed of LoS and NLoS components (depicted in Figure \ref{fig:subarray}). However, with our assumption that users are served through the LoS path, LoS interference component is expected to be the dominant one for three reasons:

\begin{enumerate}
\item The receiver's main lobe is unlikely to encompass the NLoS components of interference. The LoS component would in contrast get \textit{amplified} by the same amount as the desired signal.
 \item NLoS components are subject to higher path loss. \item NLoS components suffer from reflection loss induced by reflecting surfaces.\end{enumerate} Our simulation results for the same urban canyon scenario also validate this assumption (depicted in Figure \ref{fig:LoS_NLoS})


Therefore, we assume that intra-cell interference can be alleviated by suppressing the LoS component only. For the rest of this section, by the term interference we refer to the LoS component of intra-cell interference.

\subsection{PHY layer design: Power allocation and beamforming} \label{sec:intra_PHY}

Mitigation of co-channel interference in multiuser MIMO has been extensively studied in the literature \cite{visotsky1999optimum,wiesel2006linear,ulukus1998adaptive,rashid1998transmit,rashid1998joint}. Different approaches such as precoding, transmitter or/and receiver beamforming, power adaptation, etc. have been explored. In this section, we restrict ourselves to RF beamforming and power control to avoid the hardware complexity of digital precoders. 

In the context of power control and beamforming, there are two classical optimization problems: (a) sum-rate maximization and (b) minimum-rate maximization, subject to the power constraint(s). The former is often studied in the context of information-theoretic capacity, and does not guarantee fair sharing of resources among users. We therefore focus on the latter, which guarantees a minimum level of QoS (Quality of Service) for each of the streams. 

The minimum-rate optimization can be translated to the following problem:

\begin{align} \label{eq:opt1}
\mathcal{S}(P_T)= \begin{cases}
&\max_{\{\bomega_1,\bomega_2,\cdots,\bomega_K\}} \quad \min_{i} \, \text{SINR}_{i} \\ 
&\text{s.t.}  \phantom{\max_{\{\bomega_1,\bomega_2\}}} \sum_{k=1}^{K} \Vert \bomega_k \Vert_2^2 \leq P_T \end{cases}
\end{align}

where $\bomega_k \in \mathbb{C}^N$ is the transmit beamforming vector aimed at the k-th user, $\Vert \bomega_k \Vert_2^2$ is the power consumed by the k-th subarray, and $\text{SINR}_k$ is the signal to interference ratio at k-th receiver

\begin{align*}
\text{SINR}_k={{\left|\bomega_k^H \bf{h}_k \right|^2} \over {\sum_{\substack{i=1 \\ i \neq k}}^{K} \left|\bomega_i^H \bf{h}_k \right|^2 + \sigma_k^2} }
\end{align*}

A straightforward argument shows that (\ref{eq:opt1}) will result in the same SINR for all the users, and hence the maximum index of fairness is guaranteed.

\begin{figure*}[htb]
 \begin{minipage}[t]{0.5\linewidth}
 \centering
     {\includegraphics[width=1\textwidth]{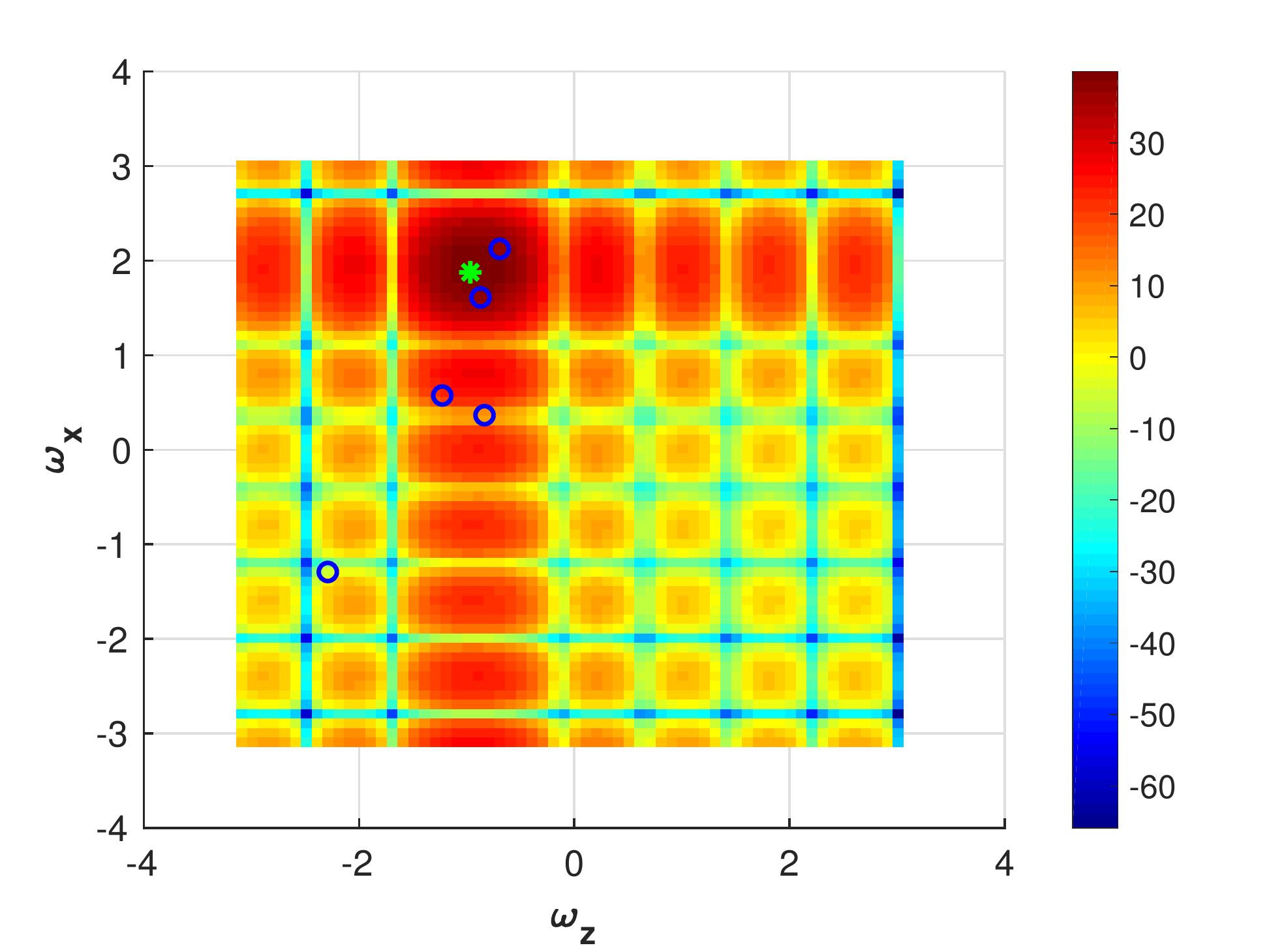}}
     \end{minipage} 
 \begin{minipage}[t]{0.5\linewidth} 
     \centering
     {\includegraphics[width=1\textwidth]{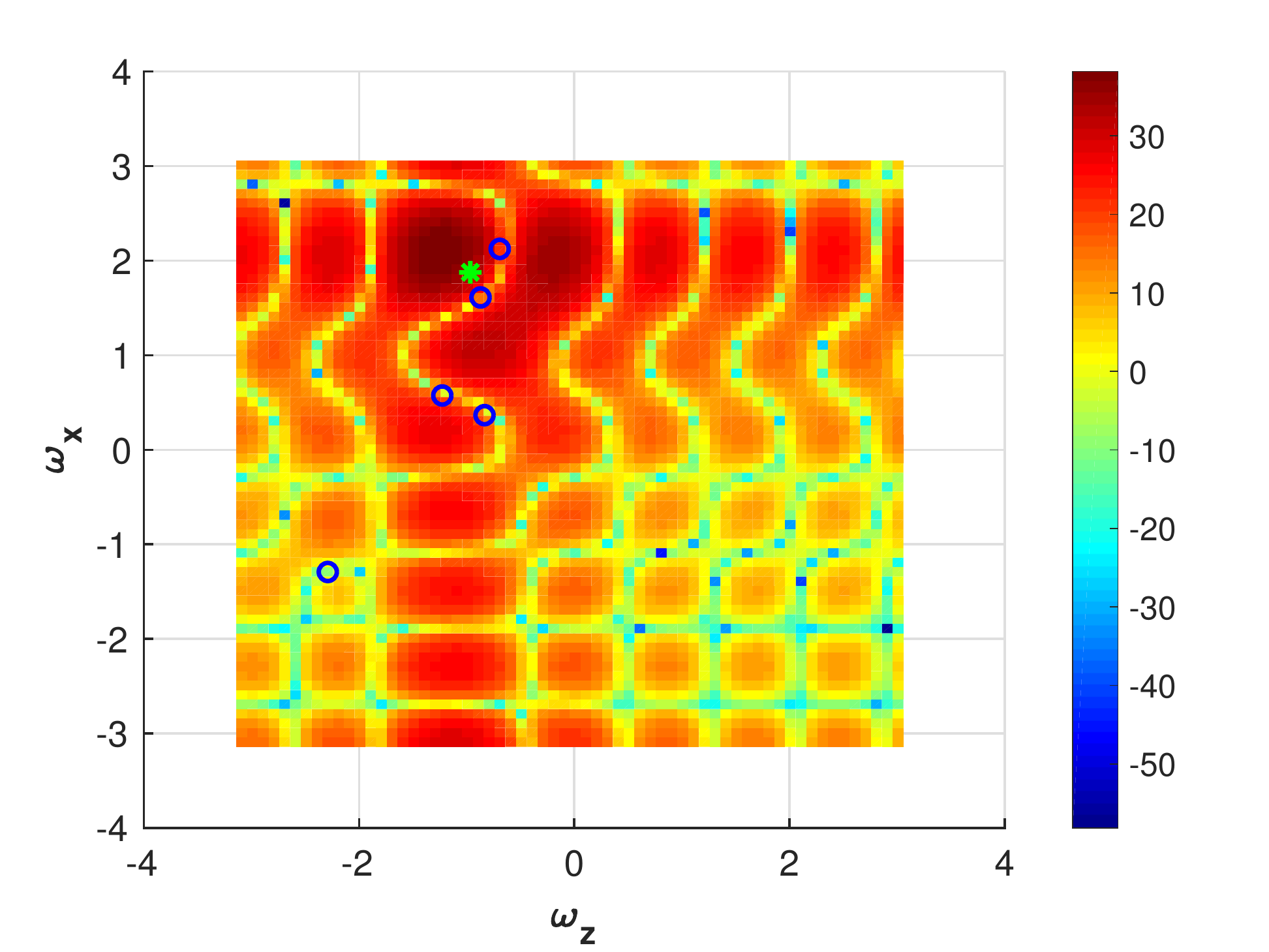}}  
 \end{minipage} 

\caption[Transmit antenna patterns before and after nullforming]{Transmit antenna patterns causing intra-cell interference (left) and the new antenna patterns after employing interference suppression via Algorithm 1 (right). The spatial frequency of the target user is marked by a green star, and the remaining users are marked by blue circles. Employing Algorithm \ref{alg:BF} (right) aligns the null directions with the non-targeted users.  
}    
 \label{fig:pattern}
 \end{figure*}

Our solution to problem (\ref{eq:opt1}) builds on previous work in \cite{visotsky1999optimum,wiesel2006linear}. We start with the related power optimization problem

\begin{align} \label{eq:opt2}
\mathcal{P}(\gamma)=\begin{cases}
&\min_{\{\bomega_1,\bomega_2,\cdots,\bomega_K\}} \sum_{k=1}^{K} \Vert \bomega_k \Vert_2^2  \\ 
&\text{s.t.} \quad \quad \quad \quad \min_{i}  \text{SINR}_{i} \geq \gamma
\end{cases}
\end{align}

It was shown in \cite{wiesel2006linear} that (\ref{eq:opt1}) and (\ref{eq:opt2}) are inverse problems, meaning that $\mathcal{S}(\mathcal{P}(\gamma_0))=\gamma_0$ and $\mathcal{P}(\mathcal{S}(P_T))=P_T$. Furthermore, (\ref{eq:opt2}) has an iterative solution \cite{visotsky1999optimum}. 
We leverage these observations to formulate Algorithm \ref{alg:BF}, which   iteratively solves (\ref{eq:opt2}) for increasing values of $\gamma$ until the power constraint in (\ref{eq:opt1}) is saturated. The solution to (\ref{eq:opt2}) employs LMMSE (Linear Minimum Mean Square Error) to estimate the transmit beamforming vector (lines 8-15) , followed by power allocation to enforce the minimum SINR constraints (line 16).

\begin{algorithm}[h]
\caption{PHY layer design} \label{alg:BF}
\begin{algorithmic}[1]
\State{\textbf{Input:}} $~~ \{p^0_i,\bh_i\} ~~ \forall i, ~~\gamma, ~~ \Delta\gamma$
\vspace{1mm}
\State{\textbf{Output:}}$~~ \{\bomega_i\} ~~ \forall i, ~~\gamma$
\vspace{1mm}
\Procedure{Beamforming and power adaptation}{}
\vspace{1mm}
\State{Compute normalized channels:} $\htilde_k = h_k/\sigma_k^2 ~~ \forall k$
\vspace{1mm}
\While{$G_{max} \Vert \bomega_k \Vert_2^2 \leq EIRP, ~~ \forall k$}
\vspace{1mm}
\State $\gamma=\gamma+\Delta\gamma$
\vspace{1mm}
\State $n \gets 0$
\vspace{1mm}
\Repeat{}
\vspace{1mm}
\For{$k \in \{1,2,\cdots,K\}$}
\vspace{2mm}
\State \parbox[t]{1\linewidth} {$\bomegahat^n_k=\argmin_{\bomega_k} \sum_{\substack{{j=1}\\{j\neq k}}}^K p_j^n |\bomega_k^H \htilde_j |^2 + \Vert \bomega_k \Vert_2^2 , \\ ~~~~~~~~~~ \text{s.t.} ~~~  \bomega_k^H \htilde_k=1$}
\vspace{3mm}
\State $p_k^{n+1}=\gamma \sum_{\substack{{j=1}\\{j\neq k}}}^K p_j^n |(\bomegahat_k^n)^H \htilde_j |^2 + \gamma \Vert \bomegahat^n_k \Vert_2^2$
\vspace{2mm}
\State $\ptilde_k^{n+1}=\gamma \sum_{\substack{{j=1}\\{j\neq k}}}^K \ptilde_j^n |(\bomegahat_j^n)^H \htilde_k |^2 + \gamma $
\EndFor
\State \textbf{end for}
\vspace{1mm}
\State $n \gets n+1$
\vspace{1mm}
\Until{ convergence}
\vspace{1mm}
\State $\bomega_k=\sqrt{\ptilde_k}\bomegahat_k^, ~~ \forall k$ 
\vspace{1mm}
\EndWhile
\State \textbf{end while}
\EndProcedure
\State \textbf{end procedure}
\end{algorithmic}
\end{algorithm}

Figure \ref{fig:pattern} illustrate how the algorithm distorts the transmitter antenna pattern by pushing nulls toward the users that the transmitter does not target.  This improves SINR but might cause SNR degradation due to sidelobe enhancement. 

\subsection*{Some remarks on the algorithm:}

\begin{itemize}
\item Intuitively, the goal of the optimization problems in (\ref{eq:opt1}) and (\ref{eq:opt2}) is to manipulate the transmitter’s antenna pattern to minimize the induced interference toward the non-targeted users while maintaining constant gain along the desired direction. In addition, power adaptation is employed to cope with link distance variations. 

\item In practice, we have individual power constraints on the Equivalent Isotropically Radiated Power (EIRP), which impose the following constraint:

\begin{align*}
G_{max} \Vert \bomega_k \Vert_2^2 \leq EIRP ~~~~ \forall k \in \{1,2,\cdots,K\}
\end{align*}

where $G_{max}$ is the maximum array gain provided by the antenna and EIRP is the limit established by FCC (Federal Communications Commission) for different frequencies (e.g., EIRP=40 dBm at 60 GHz). Our iterative solution allows us to impose the individual power constraints by setting the stopping criteria as when any of the transmit powers has reached the threshold (line 5).

\item We have omitted the effect of the receiver antenna array in our formulation.
Specifically, the channel matrix $\bH_k$ has been replaced by a vector $\bh_k$. This is for two reasons: 
\begin{enumerate}
\item In order to limit the complexity of mobile receivers, interference suppression is employed at the base station alone.
\item For intra-cell interference, the receiver antenna provides an array gain of M for both signal and interference. Thus, it does not affect performance in an interference-limited scenario.  
\end{enumerate}
However, we take the receiver arrays back into account for our simulation results (Section \ref{sec:capacity}).

\end{itemize}


\subsection{MAC layer design: Resource allocation}

The preceding PHY layer optimization is for sharing a single resource block among a pre-defined set of users. In this section, we consider interference management in the MAC layer, where resources are divided into blocks (resource granularity) and only certain users allowed to operate in each block (user selection). Intuitively, these additional degrees of freedom can be exploited in the following manner: by selecting spatially separated users to operate in the same block, we can mitigate interference and increase spectral efficiency.

\subsection*{Preliminaries:}

Consider a cell with Q users sharing frequency band B over a frame of duration T. We make two assumptions:
\begin{enumerate}
\item The frame duration T is small enough that mobile users can be considered to be quasi-stationary over a frame.
\item The directive antenna arrays employed on both transmitter and receiver suppress multipath fading sufficiently that we may approximate the channel as frequency non-selective.  
\end{enumerate} 
We consider resource allocation via time division, so that at every point in time each active user utilizes the entire bandwidth B. For simplicity we allow an infinite time granularity.

We need to allocate each time slot (small portions of a frame) to a subset of users. Denoting by $\mathcal{Q}$ the set of all users, we define $\mathcal{P}_{\leq K}(\mathcal{Q})$ as the set of all possible subsets of users (\textit{configurations}) that can be served simultanously by (up to) K antenna arrays:
$$
\mathcal{P}_{\leq K}(\mathcal{Q})=\{U_c \subset \mathcal{Q} \; \vert \; |U_c| \leq K \}
$$

We wish to find the fraction of a frame that should be allocated to each of these configurations in order to maximize sum (or minimum) spectral efficiency.
More specifically, let $x_c$ represent the portion of the time frame allocated to the c-th configuration. We want to find policy $\bx=[x_1,x_2,\cdots,x_C]^T$ where $C=\sum_{k=0}^K {Q \choose k}$ is the cardinality of $\mathcal{P}_{\leq K}(\mathcal{Q})$.

The spectral efficiency for the q-th user under policy $\bx$ is then defined as

\begin{align} \label{eq:SE}
r_q=\sum_{c=1}^C x_c \; \log(1+\gamma_{c}^q) \; \; \; \text{(bits/sec/Hz)} 
\end{align}
where $\gamma_{c}^q$ is the SINR of the q-th user under c-th configuration (where $U_c$ is the set of active users.) Clearly, we set $\gamma_{c}^q=0$, for $q \notin U_c$).

\noindent{\textbf{The resource allocation problem:} Like the optimization problems for beamforming and power adaptation, the resource allocation problem could also be formulated to maximize either the sum-rate or the min-rate. In order to provide fairness among users, we focus on the min-rate version, which can be formulated as follows: }

\begin{align} \label{eq:max-min-rate}
\max_{\bx} ~ &\min_{q} ~ r_{q} \\ \nonumber
\text{s.t.} ~~~ & S^T\bx = \mathbf{r} \\ \nonumber
& \mathbb{1}^T \bx =1 \\ \nonumber
& \bx \succeq \mathbb{0}  \nonumber
\end{align} 

In the first constraint, we have rewritten the equations in (\ref{eq:SE}) in a matrix form by defining $S_{C \times Q}=[s_{cq}]$ where $s_{cq}=\log_2(1+\gamma_{c}^q)$ is the spectral efficiency of the q-th user under the c-th configuration and $\mathbf{r}=[r_1,r_2,\cdots,r_Q]^T$ is the vector of resultant spectral efficiency over a unit time frame. The last two conditions ensure that sum of the portions allocated to different configurations add up to one and neither of them can be negative. 

Theoretically, allocation policies resulting from (\ref{eq:max-min-rate}) should maximize the min-rate among users. However, in practice we might not be able to attain the theoretical rate due to hardware constraints. If $s_M$ is the hardware-constrained spectral efficiency limit, the maximum min-rate will be bounded by ${(K/ Q)\,s_M}$. This corresponds to the \textit{saturation} point where all transmitters operate at their highest modulation rate, $s_M$. 

\begin{figure*}[ht]
 \begin{minipage}[t]{0.5\linewidth}
 \centering
     {\includegraphics[width=1\textwidth]{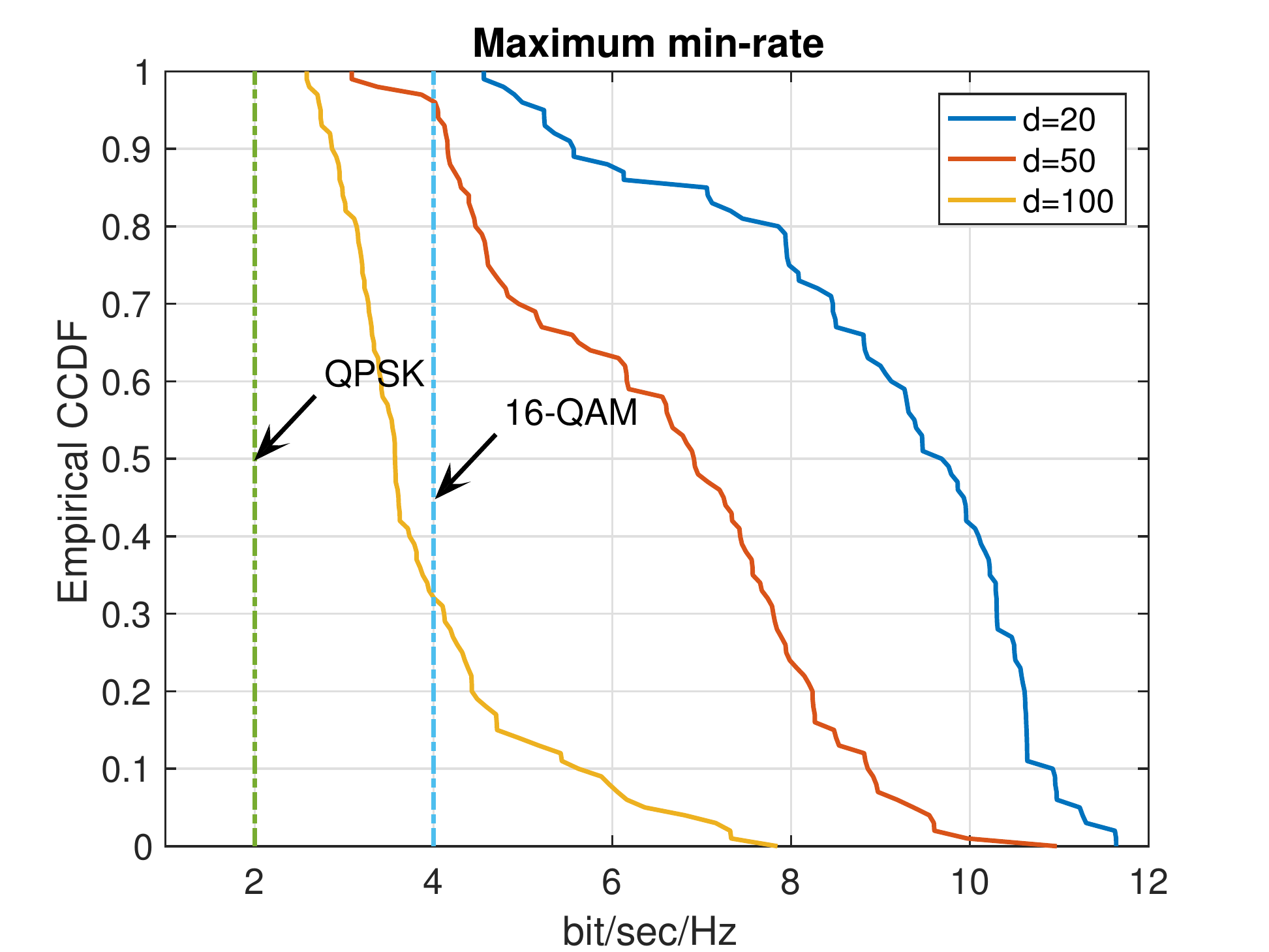}}
     \end{minipage} 
 \begin{minipage}[t]{0.5\linewidth} 
     \centering
     {\includegraphics[width=1\textwidth]{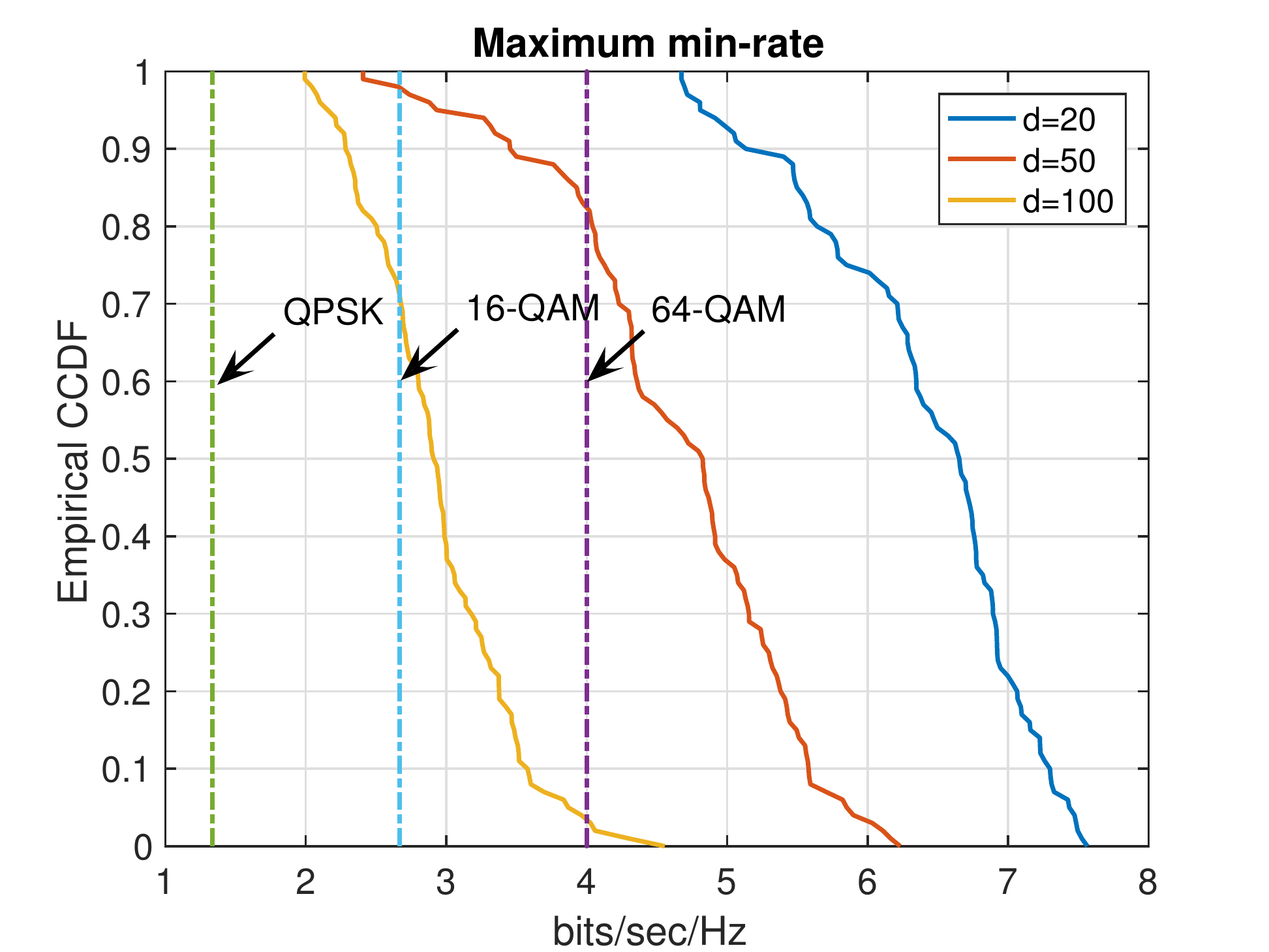}}  
 \end{minipage} 

 \caption[Empirical CCDF of the maximum min-rate]{Empirical CCDF of the maximum min-rate for (a) Q=4, K=4 (b) Q=6, K=4.}    
 \label{fig:r_g}
 \end{figure*}

Figure \ref{fig:r_g} shows the empirical CCDF of maximum min-rate for different cell sizes, along with the saturation point imposed by the various modulations (i.e., ${(K/ Q)\,s_M}$). As depicted in Figure \ref{fig:r_g}, for smaller picocells with larger number of users (d $\leq$ 20m and Q$ge$K) spectral efficiency is limited to the saturation point imposed by 64-QAM modulation ($s_M$=6 bps/Hz) and hence constrained by hardware rather than noise or interference. This is because smaller cells have (almost) vertically aligned beams which will lead to more diverse spatial frequencies as compared to less slanted beams at larger cells. As a result, our interference suppression algorithm performs more effectively in smaller cells. Furthermore, larger number of users could increase the attainable spectral efficiency by enabling us to utilize multiuser diversity to avoid interference.

\subsection*{Some remarks:}

\begin{itemize}
 
\item The optimization problem in (\ref{eq:max-min-rate}) maximizes the worst users' spectral efficiency and therefore will result in equal rate for all users in $\mathcal{Q}$.  
Its performance is therefore inherently bounded by that of the worst user. However, there are certain scenarios where we can maximize the sum-rate as well: for example, when we have surplus resources after providing all users with some minimum required spectral efficiency, $r_{min}$.

 Therefore, if the resultant min-rate provided by the allocation policy in (\ref{eq:max-min-rate}) is greater than $r_{min}$, we employ the following optimization problem to maximize the sum-rate by utilizing multiuser diversity.  

\begin{align} \label{eq:sumrate}
\max_{\bx} ~ &\mathbb{1}^T S^T\bx \\ \nonumber
\text{s.t.} ~~~ & S^T\bx \succeq r_{min}\mathbb{1} \\ \nonumber
& \mathbb{1}^T \bx =1 \\ \nonumber
& \bx \succeq \mathbb{0}  \nonumber
\end{align} 

\item An important observation is that an optimal allocation policy typically allocates more resource blocks to configurations with a larger number of users. This is because the overall datarate is linearly proportional to the number of simultaneous users, whereas the dependence on SINR is logarithmic.
However, there are settings in which time multiplexing leads to a higher data rate than spatial multiplexing (for example, when users are highly spatially correlated such that by eliminating their mutual interference, higher data rates can be attained even over smaller portion of a resource block.)

\item Figure (\ref{fig:resource_allocation}) demonstrates this phenomenon by showing a few examples for the solution to the resource allocation problem. The optimal solution tends towards serving maximum number of users simultaneously (blue portions) unless the induced interference is so large that only a subset of them are served (green or red portions). 

\begin{figure}[htb]
\centering
    {\includegraphics[width=1\columnwidth]{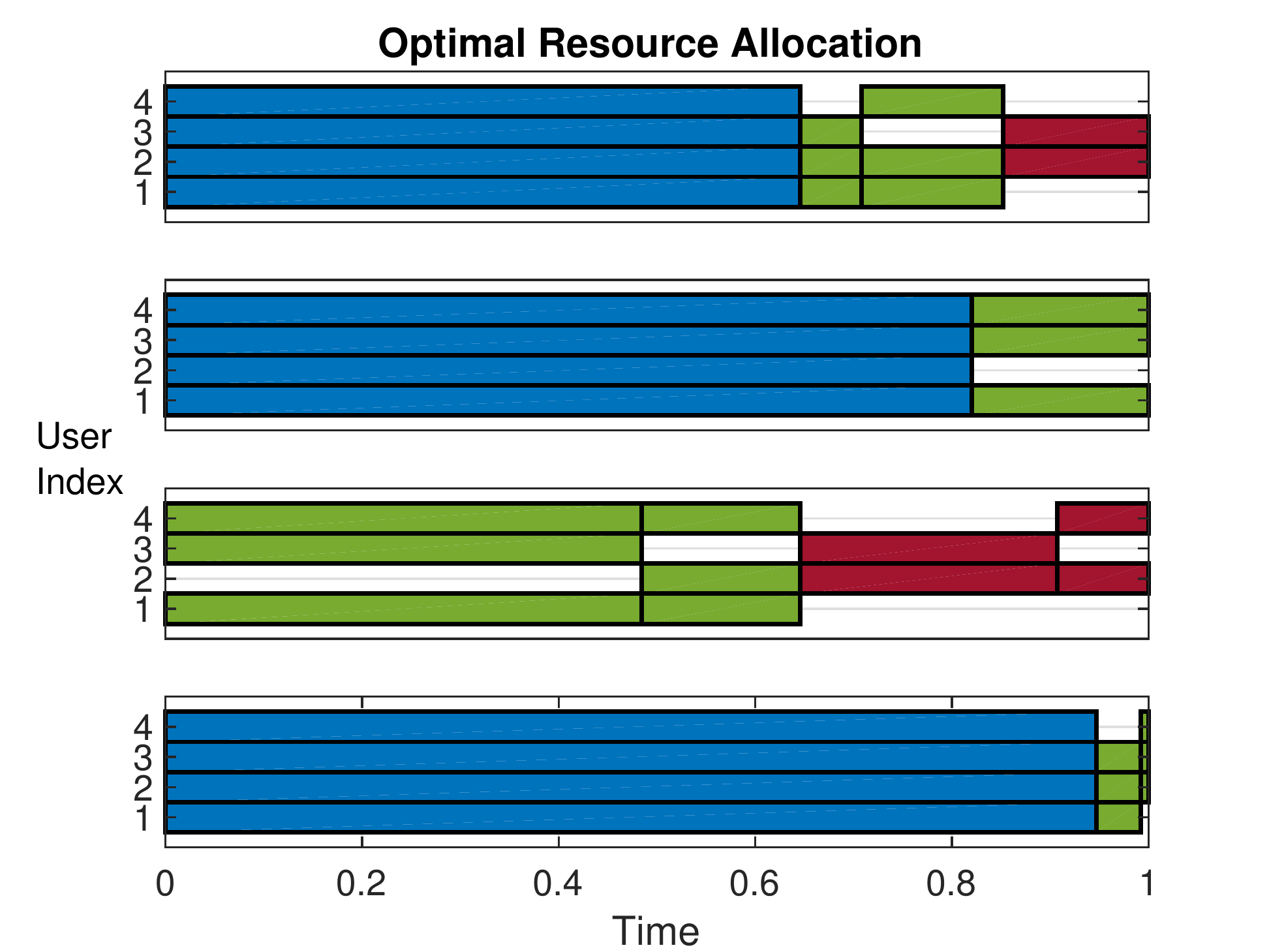}}
\caption[Optimal resource allocation]{Optimal solution of the resource allocation problem for different realizations of mobile users. The picocell parameters are d=50m and K=Q=4. Optimal allocation policies tend to serve the largest possible number of users (blue portions) while in some cases it is better to turn off a subset of subarrays, i.e., green/red portions.}   
\label{fig:resource_allocation}
\end{figure}

\end{itemize}

%% file: Capacity2.tex

We now demonstrate via simulations that mm-wave cells enjoy a significant gain in capacity over conventional LTE cellular networks, despite the increased amount of inter- and intra-cell interference. 

\subsection{Preliminaries}

\begin{figure}[tb]
\centering
     {\includegraphics[width=3.5in]{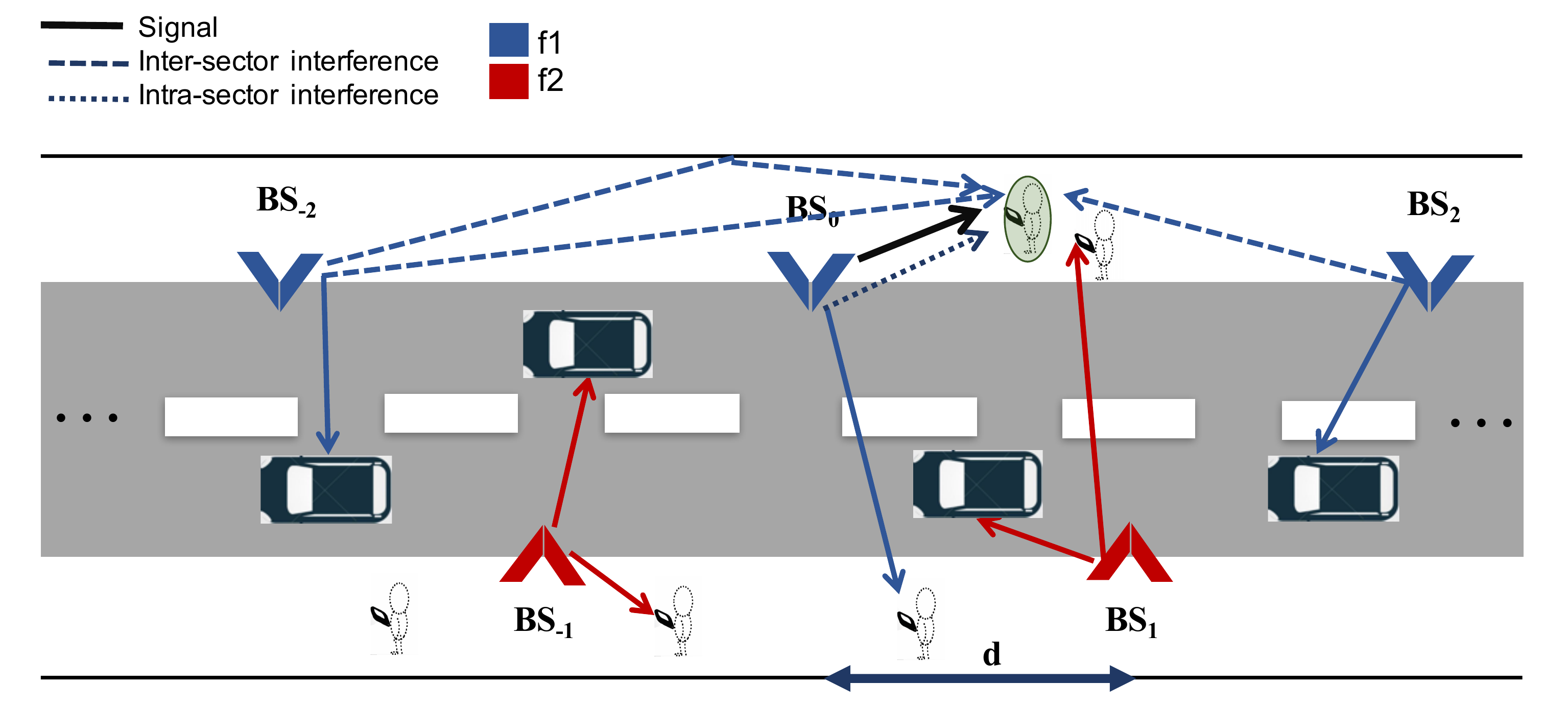}}
 \caption{Simulation scenario}   
 \label{fig:Merge}
 \end{figure}

Our interference analysis in the preceding sections is partially geometry dependent and specifically tailored for cells along an urban canyon.
Hence for our simulations, we consider an urban canyon of length 1 km and investigate a picocell in the middle of this canyon, where users would see the most interference (Figure \ref{fig:Merge}). By virtue of Theorems \ref{thm:Th1} and \ref{thm:Th2} from section \ref{sec:inter}, we ignore interference coming from outside the 1 km segment.  

Since a user in the target picocell can be served by one of two basestations on two different sides, it is unlikely for her body to block both LoS paths.  Furthermore, as we shrink the picocell width, the LoS path slants more steeply downward, hence it is difficult
for other obstacles (e.g., pedestrians, cars) to block it.  Thus, in our computations, we assume for simplicity that at least one LoS path is available
to every user.  Of course, both LoS and first order NLoS paths are accounted for when computing
interference from other subarrays. (As noted in \cite{marzi2015interference} interference from higher order reflections is negligible in comparison.)

We consider 8 $\times$ 8 basestation TX arrays and 4 $\times$ 4 mobile RX arrays. These values are chosen because they are close to the current state of the art (32 element arrays are already deployed in commercial 60 GHz products), and it turns out that they suffice to provide high spectral efficiency as we scale down cell sizes. 

By Theorems \ref{thm:Th1} and \ref{thm:Th2}, for a typical user served by BS$_0$, the interference induced by the base stations farther than 2d away from BS$_0$, is negligible. Specifically, in the scenario depicted in Figure \ref{fig:Merge}, the following sources would interfere with the shaded user served with one of the eastward facing antenna arrays of BS$_0$:

\begin{enumerate}
	\item \textit{inter-cell} interference from K eastward facing antenna arrays on BS$_{-2}$ 
	\item \textit{inter-cell} interference from K eastward facing antenna arrays on BS$_{-1}$
	\item \textit{intra-cell} interference from K-1 eastward facing antenna arrays on BS$_0$
	\item \textit{inter-cell} interference from K westward facing antenna arrays on BS$_1$
	\item \textit{inter-cell} interference from K westward facing antenna arrays on BS$_2$
\end{enumerate} 

each of which is composed of LoS and multiple NLoS components. 


In our simulations, we have employed frequency reuse of two which automatically eliminates items two and four above. We also attenuate the LoS interference of item three by employing Algorithm 1 introduced in section \ref{sec:intra_PHY}.
We compute the overall spectral efficiency, $\log_2$(1+SINR), for the users served by BS$_0$ by taking into account the residual interference from items one, three and five. The resultant matrix $S$ is then fed into the optimization problem (\ref{eq:max-min-rate}) to obtain the maximum min-rate obtained by the optimal time allocation.

\begin{figure*}[htb]
 \begin{minipage}[t]{0.5\linewidth}
 \centering
     {\includegraphics[width=1\textwidth]{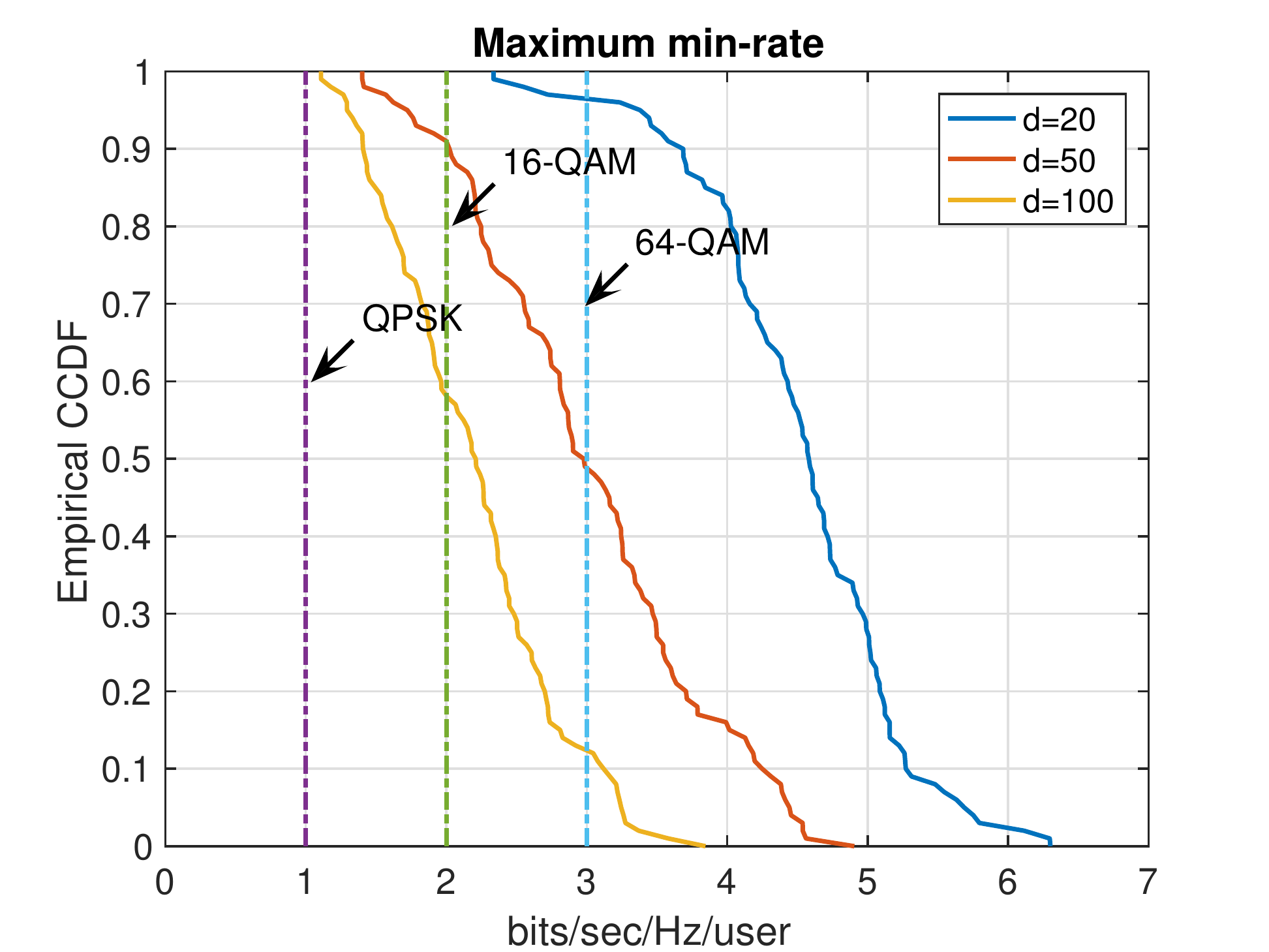}}
     \end{minipage} 
 \begin{minipage}[t]{0.5\linewidth} 
     \centering
     {\includegraphics[width=1\textwidth]{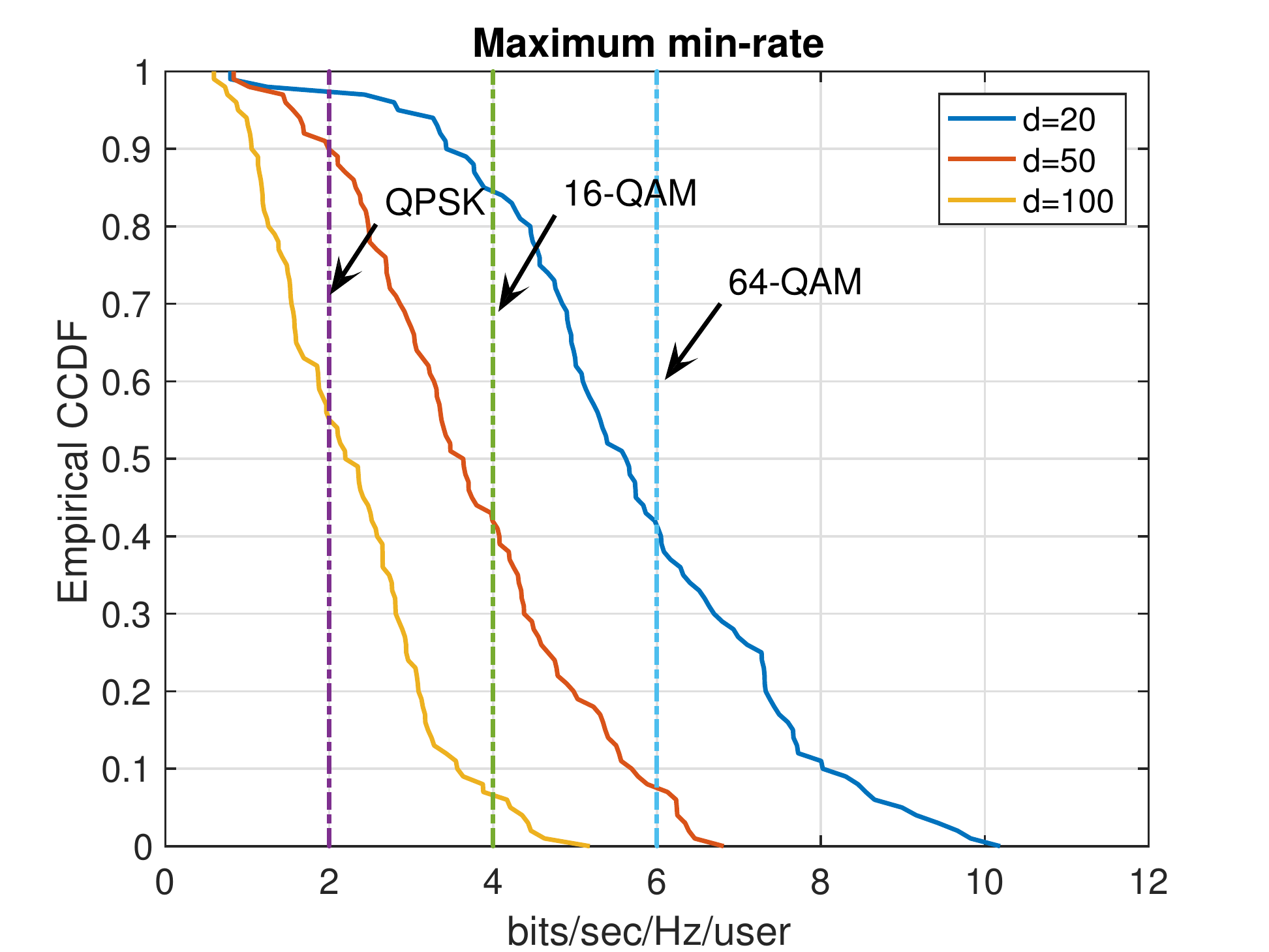}}  
 \end{minipage} 
 \caption[Empirical CCDF of the maximum min-rate]{Empirical CCDF of the maximum min-rate for users in a picocell in the middle of an urban canyon as depicted in Fig. \ref{fig:Merge}, for (a) K=2, Q=4 and (b) K=4, Q=4 .}   
 \label{fig:final_rate}
 \end{figure*}

Figure \ref{fig:final_rate} shows the empirical CCDF of the maximum min-rate provided for each of the users. Note that the hardware saturation points corresponding to QPSK, 16-QAM and 64-QAM modulations are ${(K/ Q)\,s_M}$= 1, 2 and 3 respectively for the case K=2 and Q=4. 

\subsection{Capacity calculations}

\begin{figure}[htb]
\centering
    {\includegraphics[width=1\columnwidth]{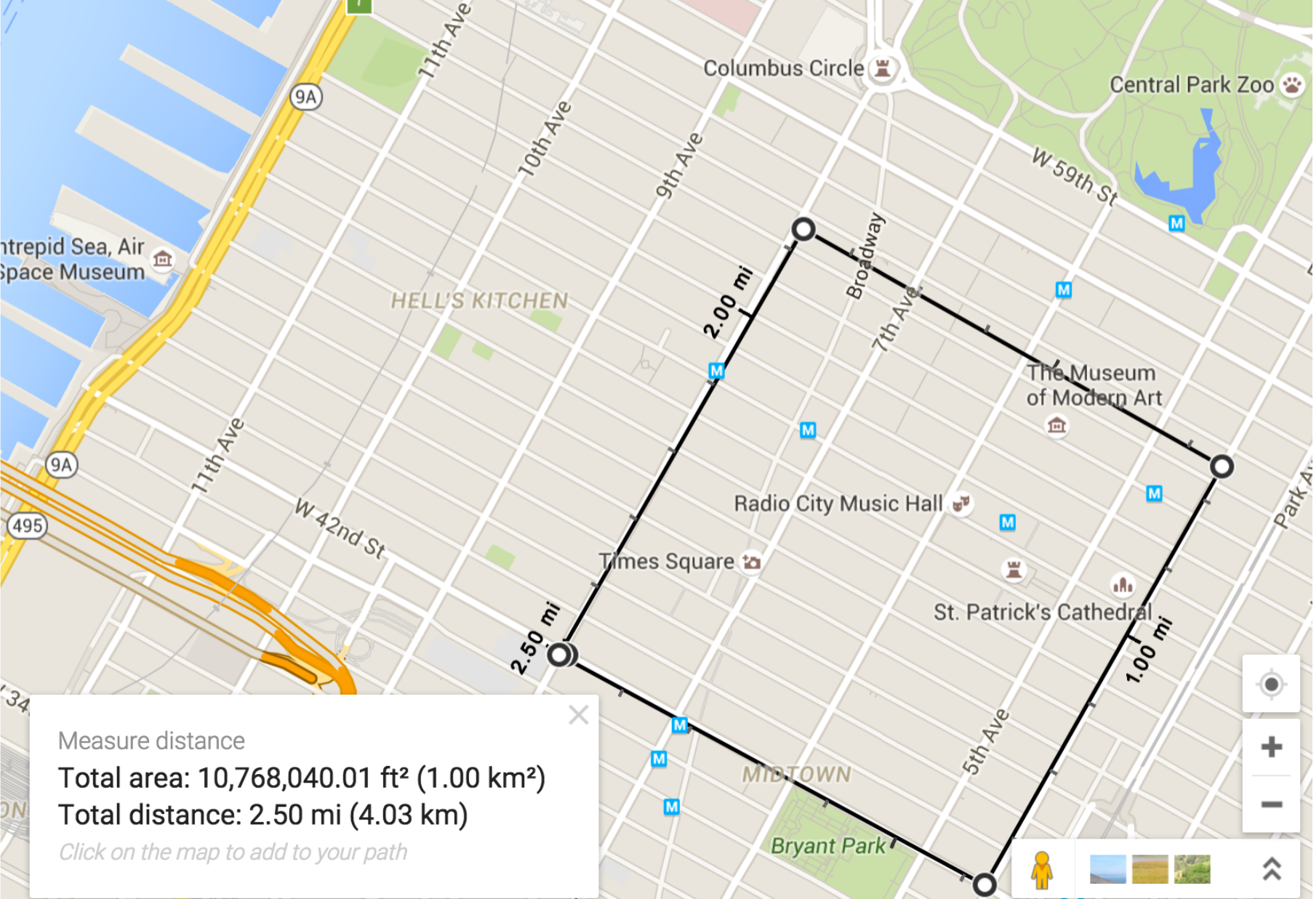}}
    \vspace{1mm}
\caption{1 km$^2$ in Manhattan area, encompassing 15 street canyons.}   
\label{fig:NY}
\end{figure}

We consider a one square kilometer region in Manhattan area (Figure \ref{fig:NY}), which encompasses 15 urban canyons.
Thus, we can get a rough estimate of the overall capacity per square kilometer of our approach via following computations:

\begin{align} \label{eq:capacity}
&\text{Capacity} \, (\text{ bps/km$^2$}) \, = \text{Maximum min-rate (bps/Hz/user)}  \times \\ \nonumber 
&\frac{B}{F} (\text{Hz})  \times 2Q \, (\text{Num. users / cell}) \times n_c \, (\text{Num. cells / km$^2$})
\end{align}

where B, F and $n_c$ are the total bandwidth, the frequency reuse factor and the number of picocells per square kilometer respectively. Note that 2Q in (\ref{eq:capacity}) refers to the number of users served within the picocell \footnote{This requires 2Q $\times$ $n_c$ = 9000 users/km$^2$ in our most extreme case: d=20m and Q=6 which is still much smaller than the population density of Manhattan area: 27,826 persons/km$^2$ \cite{wiki:Population_density}.} which are covered by either eastward facing antennas of BS$_0$ or westward facing antennas of BS$_1$. In our example of a 1km$^2$ region in Manhattan shown in Fig. \ref{fig:NY}, there are a total of 15 street canyons of length 1km (in both directions), each of which encompasses 1km$/d$ cells. Hence, we get $n_c \approx$ 150, 300 and 750 for picocell widths of d=100, 50 and 20 meters respectively. 

We have summarized the above results in table \ref{tab:Capacity} specifying the overall attainable capacity for different scenarios. Note that the Maximum min-rate in equation (\ref{eq:capacity}) is replaced with the attained rate in Fig. \ref{fig:final_rate} truncated to the hardware saturation point imposed by 64-QAM which is $(K/ Q)\,s_M$ for $s_M$=6 bps/Hz. Moreover, the first column in table \ref{tab:Capacity} corresponds to our previous results in \cite{marzi2015interference}. 

\begin{table}[htb]
\centering
\caption[Capacity (Tbps/km$^2$) for a rural area in New York]{Capacity (Tbps/km$^2$) over a total bandwidth of 2GHz for a rural area in New York employing 8 $\times$ 8 and 4 $\times$ 4 antenna arrays as transmitter and receivers.} \label{tab:Capacity}
\vspace{6pt}
\begin{tabular}{lccccc}\toprule
\multicolumn{1}{c}{Capacity (Tbps/km$^2$)} & \multicolumn{1}{c}{$K=1$} & \multicolumn{1}{c}{$K=2$} &\multicolumn{1}{c}{$K=1$} & \multicolumn{1}{c}{$K=2$} & \multicolumn{1}{c}{$K=4$}
\\
\cmidrule(l{2pt}r{2pt}){2-3} \cmidrule(l{2pt}r{2pt}){4-6}
 & \multicolumn{2}{c}{$F=1$}  &  &   \multicolumn{1}{c}{$F=2$} & \\ 
\midrule
\multicolumn{1}{c}{d=100 m} & 1.3 & 1.8 & 1.6 & 2.6 & 2.7 \\ 
\midrule
\multicolumn{1}{c}{d=50 m} & 5.3 & 8.9 & 3.3 & 6.4  & 8.9  \\ 
\midrule
\multicolumn{1}{c}{d=20 m} & 17.6 & 33.1 & 8.9 & 17.8 & 30.9 \\\bottomrule 
\end{tabular}
\end{table}

\subsection*{Some remarks:}

\begin{itemize}

\item Smaller picocells are less prone to interference since the antenna beams aiming their target users, are slanted more steeply and hence will illuminate (induce interference to) an smaller area around them. Moreover, almost vertical beams at smaller cells result in farther spatial frequencies which make it easier to isolate them with our proposed interference suppression algorithm and hence gain more from additional subarrays per face. This feature, along with the increased spatial reuse attained with smaller cell sizes, leads to massive estimated capacity of up to 30.9 Tbps/km$^2$. 

\item Larger picocells are inherently more prone to interference due to their less slanted beams, which cause severe interference to the users in a larger neighborhood around the target user. Also, they do not gain as much from more subarrays per face (Table \ref{tab:Capacity}). This is because almost horizontally aligned beams in wide cells, lead to much closer spatial frequencies for which our interference suppression algorithm is not as effective. Possible approaches to solve this problem are (a) increasing the number of antenna elements which provides more degrees of freedom for employing the interference suppression algorithm or (b) increasing basestation height which will draw users away in the spatial frequency domain.

\item Employing larger frequency reuse factor is a wasteful approach to deal with the interference for smaller cells and only lead to marginal improvement for larger cells. This was expected due to the 2X reduction in signaling bandwidth leads to a significant penalty in achievable datarates.

\end{itemize}

%% file: Conclusion_UM.tex

In this paper, we studied mm-wave picocells along an urban canyon and examined the attainable downlink capacity by employing an array of subarray architecture. We build on the inter-cell interference characterization from our previous work \cite{marzi2015interference}, and
focus here on pushing the limits of spatial reuse through cross-layer resource allocation strategies which opportunistically isolate users in the spatial and/or time domains. 

Our simulation results take both inter- and intra-cell interference into account.  We find that as we shrink the cell size (down to a cell width of 20m), 
the per-user spectral efficiency is mostly ($\geq 97\%$) bounded by hardware limitations (the bound we use is $s_M$= 6 bps/Hz,
corresponding to uncoded 64QAM). Larger cells are more prone to interference, but our proposed scheme provides users with sufficient spectral efficiency for supporting smaller constellations such as QPSK.



We now provide a rough estimate of the capacity gains attained relative to conventional LTE networks. The downlink capacity of LTE network is estimated as 0.6 Gbps/km$^2$ over a total bandwidth of 255 MHz in \cite{sauter20133g}. However, the available bandwidth for downlink cellular networks is 500 MHz, hence the total capacity could be further increased by adding more channels per base station. Therefore, we estimate the total downlink capacity of LTE networks as 1.2 Gbps/km$^2$.

\begin{table}[H]
\centering
\caption{Comparing convention LTE and mm-wave cellular networks} \label{tab:LTE}
\vspace{1mm}
\begin{tabular}{lcccc}
\toprule
 & \multicolumn{1}{c}{$\quad$LTE$\quad$} & \multicolumn{2}{c}{$\quad$mm-wave$\quad$} & \multicolumn{1}{c}{$\quad$Gain$\quad$} \\ 
\cmidrule(l){3-4}
&  & d=20 & d=100 &  \\ 
\midrule
Capacity & 1.2Gbps & 30.9 Tbps & 2.7 Tbps & $\ge$ 2250X\\
\midrule
Bandwidth & 500 MHz & \multicolumn{2}{c}{$\quad$2GHz$\quad$}  & 4X \\ 
\midrule
Spatial reuse & -- & \multicolumn{2}{c}{$\quad$--$\quad$} & $\ge$ 550X\\
\bottomrule
\end{tabular}
\end{table}

Table \ref{tab:LTE} compares the resultant capacity for mm-wave picocells computed via simulations with the benchmark capacity of LTE networks. As shown below, the targeted 1000-fold capacity increase is reachable even with the largest picocell size (d=100m) considered here. Excluding the 4X gain from the larger bandwidth of 2GHz employed in our system (which is still a small fraction of the 14GHz of available bandwidth at 60GHz), the remaining gain ($\geq$ 550X) is attained through the larger spatial reuse from small cells and pencil beams. Of course, as we have mentioned in the
introduction, many implementation challenges must be surmounted in order to attain these potential gains.  Our results provide a compelling motivation for a sustained effort in addressing these challenges.


It is worth emphasizing yet again the contrast between our results and those at lower frequencies. As we increase cell density, interference can
become a fundamental limiting factor at lower carrier frequencies \cite{ramasamy_ISIT13}. Our analysis shows that this is not the case for mm-wave frequencies: the narrow beams yield large gains in spatial reuse, which translate to orders of magnitude capacity increases.

